\documentclass[showpacs, amsmath, amssymb, amsthm, aps, jmp, thmsb, twocolumn]{revtex4}

\usepackage{color,amsbsy,amssymb,latexsym,amsfonts, amsmath,cancel}
\usepackage{mathrsfs}
\usepackage{graphicx}

\newenvironment{proof}[1][Proof]{\textbf{#1.} }{\ \rule{0.5em}{0.5em}}

\newcommand{\bel}[1]{\begin{equation}\label{#1}}                     
\newcommand{\bal}[1]{\begin{eqnarray}\label{#1}}   
\newcommand{\be}{\begin{equation}}               
\newcommand{\ba}{\begin{eqnarray}}           
\newcommand{\ee}{\end{equation}}
\newcommand{\ea}{\end{eqnarray}}

\newcommand{\braket}[1]{\left\langle #1 \right\rangle}
\newcommand{\bra}[1]{\left\langle #1 \right|}
\newcommand{\ket}[1]{\left| #1 \right\rangle}
\def\({\left(}
\def\){\right)}

\newcommand{\bea}{\begin{equation}}
\newcommand{\eea}{\end{equation}}

\usepackage{bbm}

\begin{document}
\title{Complex Probability Measure and Aharonov's Weak Value}

\author{Tatsuya Morita}
\email[Email:]{moritata@post.kek.jp}
\affiliation{Department of Physics, University of Tokyo, 7-3-1 Hongo, Bunkyo-ku, Tokyo 113-0033, Japan}

\author{Toshihiko Sasaki}
\email[Email:]{sasaki@qi.t.u-tokyo.ac.jp}
\affiliation{Photon Science Center, University of Tokyo, 7-3-1 Hongo, Bunkyo-ku, Tokyo 113-8656, Japan}

\author{Izumi Tsutsui}
\email[Email:]{izumi.tsutsui@kek.jp}
\affiliation{Theory Center, Institute of Particle and Nuclear Studies,
High Energy Accelerator Research Organization (KEK), 1-1 Oho, Tsukuba, Ibaraki 305-0801, Japan}

\date{\today}
\begin{abstract}
We present a complex probability measure relevant for double (pairs of) states in quantum mechanics, as an extension of the standard probability measure
for single states that underlies Born's statistical rule.   When the double states are treated as  the initial and final states of a quantum process, we find that  Aharonov's weak value, which has acquired a renewed interest as a novel observable quantity inherent in the process, arises as an expectation value associated with the probability measure.  Despite being complex, our measure admits the physical interpretation as mixed processes, {\it i.e.},  an ensemble of processes superposed with classical probabilities.
 
\end{abstract}

\maketitle

\paragraph*{Introduction.}

The weak value, which has long been advocated by Aharonov {\it et al.}~\cite{Aharonov_1988} as a novel and physically measurable quantity, is currently attracting much attention.   
The weak value arises as an intrinsic physical entity in the time-symmetric formulation of quantum mechanics \cite{Aharonov_1964}, where one considers a quantum process specified by both initial and final states (obtained by pre- and postselections), instead of only an initial state 
(preselection) in the usual formulation of quantum mechanics.   

There are basically two reasons for the renewed interest in the weak value.   One is that it offers an intriguing possibility of extending the scope of \lq physical reality\rq\  in quantum mechanics, and thereby sheds a new light both on paradoxical phenomena such as  Hardy's paradox \cite{Lundeen_2009, Yokota_2009}, and on fundamental issues such as the direct detection of wave function \cite{Lundeen_2011} or tracing the trajectories of Young's double slit experiment \cite{Kocsis_2011}.  
The other, which is  more important for practical purposes,  is that it serves to measure physical quantities by macroscopic apparatuses with significantly amplified precision, as has been demonstrated in the strength of interaction in the spin Hall effect \cite{Hosten_2008} or in the beam diffraction \cite{Dixon_2009}.   

Despite these impressive developments and the enhanced recognition of its significance, the weak value still remains obscure, not just because it is complex-valued and defies our direct perception as physical reality, but also because it lacks a solid basis for the uniqueness of the form, that is, why it takes the form as it is.    
The present work is an attempt to to provide such a basis from probability theory.  Specifically, 
we consider the possible form of probability measure that can be associated with pairs of states, called \lq double states\rq\ in this paper, in a manner consistent with the given states.   
Employing an argument analogous to that used in Gleason's theorem \cite{Gleason_1957} which determines the real probability measure and thereby lays a foundation of Born's statistical rule for single states, we show that the measure for double states can also be determined properly.  
Although our measure turns out to be complex in general and contains a free parameter,  it reduces in the single state limit uniquely to the conventional real measure which reproduces Born's rule.   When applied to quantum processes in which the double states are identified with the initial and final states, our complex measure is found to 
admit the interpretation of mixed processes, {\it i.e.},  ensemble of processes superposed with real probabilities, assuring its physical testability.   The weak value is then obtained as the expectation value of the observable measured in a particular ensemble corresponding to a pure process specified by the double states.  

Our attempt is not the first for realizing the weak value in the context of probability theory.   In fact, it has been argued in \cite{Steinberg_1995} that, based on the Bayesian theory, the weak value can be recognized as a conditional expectation value.   More recently, the weak value has also been realized as a conditional average over the outcomes of POVM measurements in the weak coupling limit \cite{Dressel, Hofmann_2010}.   Our approach is more direct than these, in that the conditions are encoded in the complex probability measure as an intrinsic element to the process of the system itself, where neither the introduction of conditional probability nor the consideration of weak coupling limit is necessary.   In spirit, our argument is closer to the work \cite{Hosoya_2011} which adopts a number of requirements for probability measure in different characters which are 
asymmetric in time as a whole.   In contrast, our arguments are purely based on the measure for double states  and is determined completely in a time-symmetric manner, which is more faithful to the original formulation \cite{Aharonov_1964} and seemingly much simpler.

We begin our discussion by invoking the probability measure for single states and proceed to discuss
the probability measure for double states.   We then apply our  measure to quantum processes to see how the weak value arises there, and argue that the appropriately generalized measure admits an operational interpretation as mixed processes.

\phantom{}
\paragraph*{Probability Measure for Single States.}

For our purposes, we first recall Gleason's theorem \cite{Gleason_1957} which deduces Born's statistical rule based on a probability measure fulfilling certain logical conditions.   Given a Hilbert space $\mathcal{H}$ of a finite dimension $d\geq 3$, consider a real-valued measure $\mu$ which is a map from the space of projection operators $\mathcal{P(H)}$ of $\mathcal{H}$ to non-negatives, {\it i.e.}, $\mu\colon\mathcal{P(H)}$ $\rightarrow$ $[0,\infty)\subset\mathbb{R}$.
The theorem states that, if the map is bounded $|\mu(P)| < \infty$
and satisfies the partial additivity condition,
  \begin{equation}
  \mu\left(\sum_i P_i \right) = \sum_i \mu\left( P_i \right)  ,
  \label{probcon}
  \end{equation}
for a set of projection operators $\{P_i\}$ which are mutually orthogonal $P_i P_j = \emptyset$ (null operator) for $i \ne j$, then it 
has the form,
 \begin{equation}
 \mu(P) = \mathrm{tr}(WP), 
 \label{gleason}
 \end{equation} 
with a positive self-adjoint trace class operator $W$.  

Since we have $\mu(\emptyset) = 0$ from (\ref{gleason}), if the range of the map $\mu$ is restricted to $[0,1]$,  and if the condition $\mu(\mathbbm{1}) = 1$ is further imposed, 
then obviously the map can be interpreted as a probability measure with the attached meaning that $\emptyset$ and $\mathbbm{1}$ represent propositions which are identically false and true, respectively.   Note that $\mu(\mathbbm{1}) = 1$ implies $\mathrm{tr}(W) = 1$.   

In quantum mechanics, the probability measure is indeed realized by such a measure $\mu$, where 
the operator $W$ corresponds to the density operator $\rho$ that characterizes the state of the system, and this includes the case of a Hibert space of dimension $d=2$ as well.  
For instance, if the system is described by a pure state $\rho = P_{\psi} := |\psi\rangle\langle\psi |$ for some normalized $|\psi\rangle \in \mathcal{H}$, our probability measure 
is required to yield
\begin{eqnarray}
\mu(P_{\psi}) = 1, \qquad
\mu(P_{\psi^{\perp}}) = 0.
\label{pure_state}
\end{eqnarray}
Here, the first condition states that the probability of being in the state $ |\psi\rangle$ is unity, whereas
the second states that there is no probability assigned for an arbitrary state $|\psi^{\perp}\rangle$ orthogonal to $|\psi\rangle$ for which the projection is given by $P_{\psi^{\perp}} := |\psi^{\perp}\rangle\langle\psi^{\perp} |$.   
Namely,  the measure $\mu$ has no support for the 
subspace $\mathcal{P(H_{\psi}^{\perp})} \subset \mathcal{P(H)}$, where $\mathcal{H_{\psi}^{\perp}}$ is the orthogonal complement to the one-dimensional subspace $\mathcal{H_{\psi}}  = {\rm span}\{ |\psi\rangle\}$.   We note that, because of (\ref{probcon}) the second condition in (\ref{pure_state}) actually follows from the first for a non-negative map $\mu$, but this will no longer be the case when the non-negativity is lifted.  

From (\ref{pure_state}) one finds that $W$ is uniquely determined as $W = \rho = P_{\psi}$, and this 
shows that the probability of the state $|\psi\rangle$ being in the subspace $\mathcal{H}_i \subset \mathcal{H}$ specified by the projection $P_i$ reads 
 $ \mu(P_i) = \mathrm{tr}(\rho P_i)$, which is just Born's statistical rule.   It follows that, if an observable $A$ is measured in the 
 pure state $\rho$, 
the expectation value is given by 
  \begin{equation}
 \mathcal{E}(A) := \sum_i a_i\,  \mu(P_i) = \langle\psi |A |\psi\rangle,
 \label{expreal}
   \end{equation}
where $a_i$ is an eigenvalue of $A$, and $P_i$ is the corresponding projection appearing in the spectral decomposition $A = \sum_i a_i\, P_i$.  
One notable consequence of this is that the expectation value satisfies the sum rule,
$\mathcal{E}(A + B) = \mathcal{E}(A) + \mathcal{E}(B)$ for any observables $A$, $B$ which may not commute with each other.  This implies that, although
the sum of the individual measurement outcomes of $A$ and $B$ may not be an eigenvalue of $A+B$, on average they coincide.  

\phantom{}
\paragraph*{Complex Measure for Double States.}

Now we extend the forgoing argument to a measure characterized by double states.   Let $\{\ket{\psi}, \,\ket{\phi}\}$ be two states arbitrarily chosen from $\mathcal{H}$ except that they are neither identical (up to a phase) nor orthogonal to each other ({\it i.e.}, $\bra{\phi}\psi \rangle \ne 0$).  

Analogously to the single state case (\ref{pure_state}), given the two states $\{\ket{\psi}, \,\ket{\phi}\}$ we wish to 
require
\begin{eqnarray}
\begin{array}{l}
\mu(P_\psi)=1, \qquad
\mu(P_{\psi^\perp})=0, \\
\mu(P_{\phi})= 1, \qquad
\mu(P_{\phi^\perp})=0,
 \label{pure_process_con}
\end{array}
\end{eqnarray}
where $P_{\phi} = |\phi\rangle\langle\phi |$ and
$P_{\phi^{\perp}} = |\phi^{\perp}\rangle\langle\phi^{\perp} |$ with $ |\phi^{\perp}\rangle \in \mathcal{H_{\phi}^{\perp}}$.  Obviously, in view of the uniqueness of $W$, this is impossible unless the two states are identical.  However, the condition (\ref{pure_process_con}) can be met if one promotes the measure to a complex one.

To see this, let us invoke the generalized Gleason's theorem \cite{Drisch_1979} which extends the range of the map from $[0,\infty)$ to the entire reals $\mathbb{R}$.   Demanding the condition (\ref{probcon}), one finds that such a measure $\mu_R$ admits the same form,
 \begin{equation}
 \mu_R(P) = \mathrm{tr}(W_RP),
 \label{general_Gleason}
 \end{equation} 
but  now $W_R$ is a self-adjoint trace class operator, not necessarily positive.
In order to extend the range of the map to complex numbers $\mathbb{C}$, we choose two such real maps 
$\mu_R,\, \mu'_R\colon \mathcal{P(H)} \rightarrow \mathbb{R}$ and consider
\begin{gather}
\mu_C(P) = \mu_R(P) + i\mu_R'(P),
\end{gather}
with the imaginary unit $i$.
Clearly, the map $\mu_C$ still fulfills (\ref{probcon}) by linearity and is written as
\begin{equation}
\mu_C(P) = \mathrm{tr}(W_C P) , \qquad W_C = W_R + i W_R',
\label{commes}
\end{equation}
where $W_R$ and $W_R'$ are the self-adjoint trace class operators associated with $\mu_R$ and $\mu'_R$, respectively.   
We then have:\\
{\bf Theorem}\quad {\it
If a map $\mu_C \colon \mathcal{P(H)} \rightarrow \mathbb{C}$ for $\mathcal{H}$ with finite dimension $d\geq 3$
satisfies the partial additivity condition (\ref{probcon}) and the consistency condition (\ref{pure_process_con}) for  two non-identical states $\ket{\psi}, \ket{\phi}$ with $\bra{\phi}\psi \rangle \ne 0$, then it  has the form,
\begin{eqnarray}
 \mu_C(P) &=& \mathrm{tr}(W_C P), \nonumber \\
W_C &=& \alpha\frac{|\psi\rangle\langle\phi|}{\langle\phi|\psi\rangle}
+(1-\alpha)\frac{|\phi\rangle\langle\psi|}{\langle\psi|\phi\rangle},
\label{wcresult}
\end{eqnarray}
for some $\alpha \in \mathbb{C}$.}\\
\begin{proof}
The complex measure fulfilling (\ref{probcon}) is given by (\ref{commes}) with a trace class operator $W_C$.
Let $\{|e_i\rangle; i = 1, \ldots, d \}$ be a complete orthonormal basis in $\mathcal{H}$ with $|e_1\rangle = |\psi\rangle$.   In terms of this we 
expand $W_C$ and $|\psi^{\perp}\rangle \in \mathcal{H_{\psi}^{\perp}}$ as 
\begin{equation}
W_C = \sum_{i,j = 1}^d \beta_{ij} |e_i\rangle\langle e_j |, \qquad 
|\psi^{\perp}\rangle = \sum_{i = 2}^d \gamma_i |e_i\rangle,
\end{equation} 
with $\beta_{ij}, \gamma_i \in  \mathbb{C}$.
From  (\ref{pure_process_con}) we have
\begin{eqnarray}
0 = \mu_C(P_{\psi^\perp}) &=& \sum_{i,j} \sum_{k,l\geq2} \beta_{ij} \gamma_k \gamma_l^* \mathrm{tr}( |e_i\rangle\langle e_j|e_k\rangle\langle e_l| ) \nonumber \\
 &=& \sum_{i,j\geq2} \beta_{ij} \gamma_j \gamma_i,
\end{eqnarray}
which implies $\beta_{ij} = 0$ for $i,j\geq2$ since $\gamma_i$ can be chosen arbitrarily.  
The operator $W_C$ is thus written, with some (unnormalized) states 
$|\xi_1\rangle, |\xi_2\rangle \in \mathcal{H_{\psi}^{\perp}}$, as
\begin{equation}
W_C = |\psi\rangle\langle\psi| + |\psi\rangle\langle\xi_1| + |\xi_2\rangle\langle\psi|. 
\label{compoprt}
\end{equation}
Defining
\begin{equation}
\ket{\eta} = \ket{\phi}-\braket{\psi | \phi}\ket{\psi} \in \mathcal{H_{\psi}^{\perp}},
\end{equation}
we further decompose 
\begin{equation}
|\xi_i\rangle = z_i \, |\eta\rangle + |\xi_i'\rangle, \qquad i = 1, 2,
\end{equation}
with $z_i \in \mathbb{C}$  
so that $\ket{\xi'_i} \in \mathcal{H_{\eta}^{\perp}}$ in addition to $\ket{\xi'_i} \in \mathcal{H_{\psi}^{\perp}}$.  Similarly, if we define 
\begin{equation}
|\zeta \rangle = |\psi\rangle - \langle\phi|\psi\rangle |\phi\rangle \in \mathcal{H_{\phi}^{\perp}},
\end{equation}
we find that both
$\ket{\phi}$ and $\ket{\zeta}$ belong to the linear space spanned by $\ket{\psi}$ and $\ket{\eta}$.  It follows that 
$\ket{\xi'_i} \in \mathcal{H_{\phi}^{\perp}}$ and $\ket{\xi'_i} \in \mathcal{H_{\zeta}^{\perp}}$ for $i = 1, 2$ as well.
This observation motivates us to rewrite (\ref{compoprt}) in favor of $|\phi\rangle$,  $|\zeta\rangle$ and $|\xi_i'\rangle$ to find
\begin{eqnarray}
W_C 
&=&  
\omega_{\zeta\zeta}|\zeta\rangle\langle\zeta| +
\omega_{\phi\phi}|\phi\rangle\langle\phi| + 
\omega_{\phi\zeta}|\phi\rangle\langle\zeta| +
\omega_{\zeta\phi}|\zeta\rangle\langle\phi| 
\nonumber \\
&&\!\!\!\!\!\!\!\!\!\!\! +\, \omega_{\phi\xi_1}|\phi\rangle\langle\xi_1'| + \omega_{\xi_2\phi}|\xi_2'\rangle\langle\phi|
+ |\zeta\rangle\langle\xi_1'| + |\xi_2'\rangle\langle\zeta|,
\label{wcone}
\end{eqnarray}
where
\begin{eqnarray}
\begin{array}{l}
\omega_{\zeta\zeta} = 1-z_1^*\langle\phi|\psi\rangle-z_2\langle\psi|\phi\rangle,  \\
\omega_{\phi\phi} =  1 - \omega_{\zeta\zeta} + \omega_{\zeta\zeta}|\langle\psi|\phi\rangle|^2 ,  \\
\omega_{\phi\zeta} =  z_2   +\omega_{\zeta\zeta}\langle\phi|\psi\rangle, \\
\omega_{\zeta\phi} =  z_1^*   +\omega_{\zeta\zeta}\langle\psi|\phi\rangle, \\
\omega_{\phi\xi_1} =  \langle\phi|\psi\rangle, \qquad \omega_{\xi_2\phi} = \langle\psi|\phi\rangle.
\label{setofsoln}
\end{array}
\end{eqnarray}

On the other hand,  an analogous argument  for the state $|\phi\rangle$ demanded by (\ref{pure_process_con}) shows that $W_C$ must also be of the form, 
\begin{equation}
W_C = |\phi\rangle\langle\phi| + |\phi\rangle\langle\chi_1| + |\chi_2\rangle\langle\phi|, 
\label{wctwo}
\end{equation}
with  some (unnormalized) states $|\chi_1\rangle, |\chi_2\rangle \in \mathcal{H_{\phi}^{\perp}}$.
Since $\ket{\psi}$ and $\ket{\phi}$ are not identical, we have $\ket{\zeta} \ne 0$.
Comparing (\ref{wcone}) and (\ref{wctwo}), we obtain
\begin{gather}
\omega_{\zeta\zeta} = 0 \label{xycon}, \qquad |\xi_1'\rangle = |\xi_2'\rangle = 0,
\end{gather}
which implies
\begin{eqnarray}
W_C &=& |\psi\rangle \langle\psi| + z_1^*|\psi\rangle \langle\eta| + z_2|\eta\rangle\langle\psi| \nonumber \\
 &=& z_1^*|\psi\rangle \langle\phi| + z_2 |\phi\rangle\langle\psi|. 
\end{eqnarray}
Since $z_1$, $z_2$ are free parameters  but subject to $\omega_{\zeta\zeta} = 0$, 
we arrive at (\ref{wcresult}) after putting $\alpha = z_1^* \langle\phi|\psi\rangle$. 
\end{proof}

Having found the complex measure $\mu_C$ for double states specified by (\ref{pure_process_con}), we may consider
the expectation value of an observable $A$.
Putting aside the question of the meaning of complex probability for the moment, and assuming if necessary that  (\ref{pure_process_con}) is valid also for the case of dimension $d=2$,
we just follow the standard 
construction of the expectation value as we did in (\ref{expreal}) to find
\begin{eqnarray}
\lambda(A) &:=& \sum_i a_i \, \mu_C(P_i) \nonumber \\
&=& \alpha\frac{\langle \phi |A|\psi\rangle }{\langle\phi|\psi\rangle } +
(1-\alpha)\frac{\langle \psi |A| \phi \rangle }{\langle\psi|\phi\rangle }. 
\label{complex_prob}
\end{eqnarray}
We then notice that Aharonov's weak value $A_w = {\langle \phi |A|\psi\rangle }/{\langle\phi|\psi\rangle}$ arises at the choice $\alpha = 1$ of the expectation value $\lambda(A)$.  
Although $\lambda(A)$ is complex in general, 
it becomes real at  
$\alpha = 1/2$ where $W_C$ becomes self-adjoint.  This shows that one can find the measure $\mu$ 
which meets the condition (\ref{pure_process_con}) by extending the range of the map only to the entire $\mathbb{R}$, but we shall soon realize that the particular measure 
obtained by $\alpha = 1/2$ does not account for all possible cases when applied to quantum processes.
Notice also that, as for $\mathcal{E}(A)$ the sum rule holds, 
 $\lambda(A + B) = \lambda(A) + \lambda(B)$, for any $\alpha$.  
 
 Interestingly, in the single state limit, that is, in the limit $|\phi\rangle \to |\psi\rangle$ the ambiguity of $\alpha$ disappears and our complex measure $\mu_C$ reduces to the 
 real measure $\mu$ in (\ref{gleason}) with the condition (\ref{pure_state}) enforced.  Accordingly, the expectation value $\lambda(A)$ also reduces to the conventional one 
 $\mathcal{E}(A)$ in (\ref{expreal}).   
 
Another observation worth mentioning is that, since under the single state $|\psi\rangle$
the probability $\mu(P_\phi) = \vert \langle\psi|\phi\rangle \vert^2$ represents the compatibility of the double states $|\phi\rangle$ and $|\psi\rangle$, one obtains the overall expectation value of $A$ by the weighted product,
$\vert \langle\psi|\phi\rangle \vert^2 \lambda(A)$.   
The average value obtained after allowing the state $|\phi\rangle$ to vary freely may then be evaluated by 
\begin{eqnarray}
\sum_{|\phi\rangle \in \mathcal{B}} \vert \langle\psi|\phi\rangle \vert^2 \lambda(A) = \langle\psi |A |\psi\rangle,
\label{aaverage}
\end{eqnarray}
where the summation is over the states of a complete basis $\mathcal{B}$ of $\mathcal{H}$.  
This outcome (\ref{aaverage}) implies that the average is actually independent of the choice of the basis set $\mathcal{B}$.   This contextual independence has been mentioned earlier \cite{Hosoya_2010, Hosoya_2011}, but here we notice that it is also the case even with the parameter $\alpha$.
Similarly, if one averages over $|\psi\rangle$ instead, one obtains $\langle\phi |A |\phi\rangle$.
In other words, the information of one of the double states will disappear once the average of the state is taken, and the outcome becomes the standard expectation value with respect to the other state left over.

So far, we have considered complex measures with $W_C$ of the type (\ref{wcresult}) which fulfils (\ref{pure_process_con}).  As one can extend the scope of single states from pure states to mixed states by allowing $W$ to be any positive self-adjoint operators with unit trace, one may similarly extend the scope of double states by allowing $W_C$ to be any operators with unit trace.    
If we let $\mathcal{T(H)}$ be the space of operators with unit trace, we have
\begin{eqnarray}
\beta\, W_C + (1-\beta)\, W_C' \in \mathcal{T(H)},
\end{eqnarray}
for $W_C, W_C' \in \mathcal{T(H)}$ and $\beta \in \mathbb{C}$.   This shows that, if we regard 
$\mathcal{T(H)}$ as the space of such generalized double states, the space is \lq convex\rq\ in the complex sense.  
The measure $\mu_C(P)$ also provides the map 
${\rm Pr}(P; W_C) := \mu_C(P) = \mathrm{tr}(W_C P)$,  which enjoys the affine property,
\begin{eqnarray}
&&\!\!\!{\rm Pr}(P; \beta\, W_C + (1-\beta)\, W_C')  \nonumber \\
&&\quad = \beta\, {\rm Pr}(P; W_C) + (1-\beta)\, {\rm Pr}(P; W_C'),
\end{eqnarray}
analogous to the conventional probability map.   Along with the property ${\rm Pr}(\emptyset; W_C) = 0$ and ${\rm Pr}(\mathbbm{1}; W_C) = 1$, 
this may be considered as a formal support for  $\mu_C(P)$ qualifying as a probability measure, albeit it is complex.
To explore the possible use of the complex measure, and thereby examine the physical significance of the complex parameter $\alpha$
in (\ref{wcresult}) or (\ref{complex_prob}), we now turn to
the probability measure for a quantum process.

\phantom{}
\paragraph*{Probability Measure for a Process.}

The complex measure for double states can be used to furnish the probability measure for a quantum process $\ket{\psi} \to \ket{\phi}$ by taking 
the time dependence of the states properly into consideration.  
Let $t_i$ and $t_f$ be the initial time and the final time at which the states $\ket{\psi}$ and $\ket{\phi}$ are realized, respectively, and let $t$ be the time of 
\lq measuring\rq\ the obserbable $A$ in the period, $t_i \leq t \leq t_f$.   To evaluate the outcome of the measurement results, 
we would like to have the complex measure relevant at time $t$.  
Assuming that our system is closed during the process, we have a unitary operator $U$ to describe the time development in the period.   
The forward time-developed state at $t$ from the initial state is then given by $U(t-t_i)\ket{\psi}$, and the backward time-developed state at $t$ from the final state is
given by $U(t-t_f)\ket{\phi}$.   This suggests that, instead of the two states $\ket{\psi}$, $\ket{\phi}$, we should use these forward and backward time-developed states
to characterize the measure (\ref{wcresult}).   We are thus led to 
\begin{eqnarray}
&&\!\!\!\!\!\!W_C(t) =  \alpha\frac{U(t-t_i)\ket{\psi}\bra{\phi} U^\dagger(t-t_f)}{\bra{\phi}U(t_f-t_i)\ket{\psi}}  \nonumber  \\
 && \quad +\,  (1-\alpha)\frac{U(t-t_f)\ket{\phi}\bra{\psi}U^\dagger(t-t_i)}{\bra{\psi}U^\dagger(t_f-t_i)\ket{\phi}},
 \label{timeexpw}
\end{eqnarray}
from which we can obtain the time-dependent measure,
$ \mu_C(P; t) := \mathrm{tr}(W_C(t) P)$.  The expectation value (\ref{complex_prob}) then acquires the corresponding time-dependence by the use of 
$\mu_C(P;t)$, which is characterized by the consistency condition at the initial and final times,
\begin{eqnarray}
\begin{array}{l}
\mu_C(P_\psi; t_i)=1, \qquad
\mu_C(P_{\psi^\perp}; t_i)=0, \\
\mu_C(P_{\phi}; t_f)= 1, \qquad
\mu_C(P_{\phi^\perp}; t_f)=0.
 \label{pure_process_contwo}
\end{array}
\end{eqnarray}

\begin{figure}[t]
 \begin{center}
\includegraphics[width=55mm]{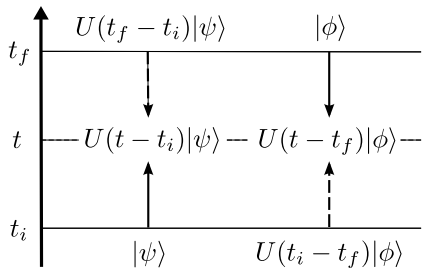}
 \end{center}
\caption{The double state $W_C(t)$ in (\ref{timeexpw}) can equally be associated with either of the two pure processes, $\ket{\psi}\rightarrow\ket{\phi}$ shown by the solid arrows or its dual $U(t_i-t_f)\ket{\phi}\rightarrow U(t_f-t_i)\ket{\psi}$ shown by the dashed arrows, representing the forward and backward time developments.
 }
\label{fig:two_process}
 \end{figure}

However, the identification of the operator $W_C$ by (\ref{timeexpw}) with the process $\ket{\psi} \to \ket{\phi}$ in the period $[t_i, t_f]$ is not quite correct, because
our measure for double states is originally given at a single time $t$ and does not involve the time direction in any intrinsic manner.  
In fact, one can also associate the same $W_C$ in (\ref{timeexpw}) with the \lq dual\rq\ process
$U(t_i-t_f)\ket{\phi} \to U(t_f-t_i)\ket{\psi}$ 
in the same period, since the two states that determine the double states at time $t$ are equivalent in both cases (see Fig.~\ref{fig:two_process}).  
In fact, with $\tilde{\ket{\phi}} := U(t_i-t_f)\ket{\phi}$, $\tilde{\ket{\psi}} := U(t_f-t_i)\ket{\psi}$, one can equally characterize our measure by
\begin{eqnarray}
\begin{array}{l}
\mu_C(P_{\tilde\phi}; t_i)=1, \qquad
\mu_C(P_{{\tilde\phi}^\perp}; t_i)=0, \\
\mu_C(P_{\tilde\psi}; t_f)= 1, \qquad
\mu_C(P_{{\tilde\psi}^\perp}; t_f)=0,
 \label{pure_process_conthree}
\end{array}
\end{eqnarray}
instead of (\ref{pure_process_contwo}).   
This indicates that the proper interpretation of $W_C$ is that it is the measure corresponding to a linear superposition of the two processes, with the parameter of the 
superposition $\alpha$.   In particular, the choice $\alpha = 1$ yields the pure process $\ket{\psi} \to \ket{\phi}$ whereas the choice $\alpha = 0$ yields another pure
process $\tilde{\ket{\phi}}  \to \tilde{\ket{\psi}}$.   Through this identification we can establish a one-to-one correspondence between the measure and the superposed process, and thereby remove the ambiguity in the expectation value $\lambda(A)$.   As a result,  $\lambda(A)$ agrees precisely with Aharonov's weak value $A_w$ for $ W_C(t)$ when the process is $\ket{\psi} \to \ket{\phi}$.

On physical grounds, one may argue that this interpretation is marred by the complex nature of the superposition.    This concern can be dismissed by invoking the fact that
any operator $W_C \in \mathcal{T(H)}$ admits the expansion,
\begin{eqnarray}
W_C = \sum_{i=1}^{d+1} p_i \, \frac{\ket{\psi_i}\bra{\phi_i}}{\bra{\phi_i} \! \psi_i\rangle}, \qquad p_i \in [0, 1],
\label{moritaexp}
\end{eqnarray}
in terms of pairs of states $\ket{\psi_i}$, $\ket{\phi_i} \in \mathcal{H}$ for which $\bra{\phi_i}\! \psi_i\rangle \ne 0$ for $i = 1, \ldots, d+1$ (see the Appendix).
Since $ \mathrm{tr}(W_C) = \sum_i p_i = 1$, one can interpret the (generalized) double states as a superposition of processes $\ket{\psi_i} \to \ket{\phi_i}$
with the real probability $p_i$ (hereafter we omit the time dependence for simplicity, pretending $U = \mathbbm{1}$).
One therefore sees that, since the complex superposition of processes can actually be viewed as a classical probability mixture of different processes.
We may call those double states with $W_C^2 = W_C$ \lq pure\rq, and otherwise call them \lq mixed\rq, knowing that the former has only one term in the expansion (\ref{moritaexp}) and reads
$W_C = \ket{\psi}\bra{\phi}/\bra{\phi} \! \psi\rangle$ for some pair of states, $\ket{\psi}$ and $\ket{\phi}$.
This interpretation provides the physical meaning of the parameter $\alpha$ in (\ref{timeexpw}).  

We may illustrate the expansion (\ref{moritaexp}) for the type of double states given in (\ref{wcresult}) by a system of a spin-1 particle.
Let our complex measure $W_C$ be determined by the state polarized up along x-direction $\ket{S_x=+1}=\ket{x_+}$ and the state polarized up along z-direction $\ket{S_z=+1}=\ket{z_+}$ as
\begin{equation}
W_C = \alpha\frac{\ket{x_+}\bra{z_+}}{\braket{z_+|x_+}} + \(1-\alpha\)\frac{\ket{z_+}\bra{x_+}}{\braket{x_+|z_+}}. 
\label{xzexample}
\end{equation} 
According to (\ref{moritaexp}), $W_C$ can be expanded into four possible processes with real probabilities $\{p_i\}$ which can be chosen arbitrarily.   
For instance,  one may assign $p_i = 1/4$ equally to each of the four processes $\ket{\psi_i} \to \ket{\phi_i}$ given by
\begin{eqnarray}
\begin{array}{l}
\!\!\ket{\psi_1} =  \ket{z_+}\! , \,\,  \ket{\phi_1} = (1-\alpha^*)(\sqrt{2}\ket{z_0}+\ket{z_-}) + \frac{\sqrt{3}}{4}\ket{\omega},  \\
\!\!\ket{\psi_2} =  \ket{z_0}\! ,  \,\,\, \ket{\phi_2} = (1+\sqrt{2}\alpha^*)\ket{z_+}-\frac{\sqrt{3}}{4}\ket{\omega}\! , \\
\!\!\ket{\psi_3} =  \ket{z_-}\! , \,\, \ket{\phi_3} =  (1+\alpha^*)\ket{z_+}-\frac{\sqrt{3}}{4}\ket{\omega}\! , \\
\!\!\ket{\psi_4} =  \ket{\omega}\! , \,\,\,\,\, \ket{\phi_4} =  \ket{z_+}-\frac{\sqrt{3}}{4}\ket{\omega}\! , 
\label{decomstate}
\end{array}
\end{eqnarray}
up to normalization, with $\ket{\omega}=\(\ket{z_+}-\ket{z_0}-\ket{z_-}\)/\sqrt{3}$.   This shows that, once the probabilities $\{p_i\}$ are chosen,  the parameter $\alpha$ determines 
the corresponding physical processes into which the process of the double states (\ref{xzexample}) is decomposed.  
We emphasize that, since the expansion (\ref{moritaexp}) is highly non-unique, so is the set of processes decomposed.  In fact, even if one has $\alpha\in[0,1]$ so that the double state (\ref{xzexample}) can be directly interpreted as a probability mixture of the two processes 
$\ket{z_+} \to \ket{x_+}$ and $\ket{x_+} \to \ket{z_+}$ with $p_1 = \alpha$ and $p_2 = 1 - \alpha$, one may still seek other modes of decomposition for the same double state (\ref{xzexample}).   The one (\ref{decomstate}) presented above is just one example.

The probabilistic interpretation admitted by the expansion (\ref{moritaexp}) explains how the expectation value of the observable $A$ can be obtained.  Indeed, if we consider
the measurement of $A$ in an ensemble of processes $\ket{\psi_i} \to \ket{\phi_i}$ each of which prepared with probability $p_i$, the average of the weak values reads
\begin{eqnarray}
\sum_{i} p_i \, \frac{\bra{\phi_i} \!A\! \ket{\psi_i}}{\bra{\phi_i} \! \psi_i\rangle} =  \mathrm{tr}(W_C A),
\label{valuemoritaexp}
\end{eqnarray}
which agrees with $\lambda(A)$ in (\ref{complex_prob}) for the case (\ref{wcresult}).

Our double states $W_C$ in the general context of quantum processes are closely related to the \lq two-states\rq\ proposed earlier in \cite{Aharonov_1995} which are defined by $W_T = \sum_{i, j} c_{ij} \ket{e_i}\bra{f_j}$ with $c_{ij} \in \mathbb{C}$.  The states appearing in the expansion are from the bases $\mathcal{B}_e = \{\ket{e_i} \}$, $\mathcal{B}_f = \{\ket{f_j} \}$ in $ \mathcal{H}$ which are mutually non-orthogonal  $\bra{e_i}\! f_j\rangle \ne 0$ for any $\ket{e_i} \in \mathcal{B}_e$, $\ket{f_j} \in \mathcal{B}_f$.   Due to this non-orthogonality condition, these $W_T$, if normalized, form a subspace of the space of double states $\mathcal{T(H)}$.   The difference looks subtle but conceptually it is not,  since  
the two-state $W_T$ represents processes of an open system which is realized as a subsystem of an extended closed system \cite{Aharonov_1995, Aharonov_2009}.  In contrast, our $W_C$ is given a direct operational interpretation of a probability ensemble of processes with no need of extension, and it represents the most general situations realized by such an ensemble in quantum mechanics.

\phantom{}
\paragraph*{Conclusion and Discussions.}

In this paper we have considered the complex probability measure for double states as an extension of the real probability measure for single states.  Our intention is to provide an appropriate framework for discussing probabilities and physical quantities in a given circumstance specified by two conditions, which are basically determined by two states such as the initial and final states prescribing a quantum process. 
In order to find out the form of the measure, we have employed an analogous argument used for Gleason's theorem  which establishes Born's rule of statistical interpretation for single states, implementing the complex extension and the additional condition required by the double states.  
The complex measure $\mu_C$ turns out to be rather simple (\ref{wcresult}) and contains a free parameter $\alpha$.   With this measure we have evaluated the expectation value $\lambda(A)$ of an observable $A$ and found that it coincides with Aharonov's weak value at $\alpha = 1$.   
The standard mathematical scheme for describing probabilities associated with quantum states is provided by noncommutative measure theory based on von Neumann algebras, where 
the required measures are realized by single states (called \lq normal states\rq).   Our probability theory of quantum processes can also be put on a similar basis, where the required measures are now complex and realized by double states. 

In the application of the measure for describing quantum processes, we are alluded to the interpretation that
the measure represents a superposition of two processes with the weight specified by $\alpha$.  The case $\alpha = 1$ is then found to be one of the two exceptional situations of realizing a pure process (the other being $\alpha = 0$), in accordance with the original interpretation of the weak value \cite{Aharonov_1988}.   We also have extended our measure to accommodate the most general case of superpositions, in parallel with the case of single states.   Our generalized double states admit a direct operational interpretation in terms of classical probability ensembles of processes, which is important to assure that the complex measure is meaningful in describing actual physical situations.   

We mention that the notion of complex probability in quantum mechanics appeared a long time ago in the expansion of density states in terms of two complete basis states \cite{Kirkwood}, where the coefficients are interpreted as complex joint probability (see also \cite{Hofmann_2012}).   In fact, this may be thought of just as our complex measure with positive $W_C$ in such decomposition.   The double states $W_C$ of  (\ref{wcresult}) at $\alpha = 1/2$  also appear in the weak measurement tomography as the transient density matrix \cite{Hofmann_2010}.  These instances indicate that our complex probability measure based on the double states may have more applications than the one mentioned here.

In our probabilistic treatment, the weak value $A_w$ is obtained as the expectation value $\lambda(A)$.  However, it has been argued that $A_w$ should actually be regarded as an intrinsic value possessed by the system in the given process, which can be confirmed by measuring it without (almost) disturbing the system, {\it i.e.}, by weak measurement. 
The gap between the two realizations of $A_w$, as an expectation value and an intrinsic value, seems to suggest that a deeper understanding of the value is still in need.

\begin{acknowledgments}
We appreciate Professors Y. Aharonov, H. F. Hofmann, A. Hosoya, M. Iinuma, N. Imoto, Y. Shikano and A. Wipf for useful discussions.
\end{acknowledgments}

\appendix
\section*{Appendix}

In this Appendix we show that any double state $W_C$ can be expanded as ({\ref{moritaexp}}) in terms of  
states $\ket{\psi_i}, \ket{\phi_i} \in \mathcal{H}$ for which $\bra{\phi_i}\! \psi_i\rangle \ne 0$
under the probabilities $p_i \in [0, 1]$ with $\Sigma_{i=1}^{d+1} p_i =1$.   We do this by furnishing 
the states explicitly as follows.

Let $\{\ket{w_i};i=1,...,d\}$ be a complete orthogonal basis $\mathcal{B}$ of $\mathcal{H}$.
A double state $W_C$ is then expanded as 
\begin{eqnarray}
W_C = \sum_{i, j =1}^d w_{ij} \ket{w_i}\bra{w_j},
\label{wexpansion}
\end{eqnarray}
with $w_{ij}  \in \mathbb{C}$ and $\sum_i w_{ii} = 1$, and $w_{1k}\not=0$ for some $k$.
Consider then the states $\ket{\psi_i}, \ket{\phi_i}$ for $i=1,\ldots, d$ given by
\begin{eqnarray}
\ket{\psi_i} &=& \ket{w_i}, \nonumber\\
\ket{\phi_i} &=& \sum_{j=1,j\not=i}^d \beta_{ij}^*\ket{w_j} + \ket{w_i},
\end{eqnarray}
\newline
with $\beta_{ij} \in \mathbb{C}$ and also for $i=d+1$ by
\begin{eqnarray}
\ket{\psi_{d+1}} &=& \sum_{i=2}^{d}\alpha_i\ket{w_i} + \ket{w_1},\nonumber\\
\ket{\phi_{d+1}} &=& \sum_{i=2}^{d}\ket{w_i} + \Bigl(1-\sum_{j=2}^{d}\alpha_j^*\Bigr)\ket{w_1},
\end{eqnarray}
with $\alpha_i \in \mathbb{C}$.   Note that these pairs of states fulfill $\bra{\phi_i}\! \psi_i\rangle = 1$ for all $i = 1, \ldots, d+1$. 
Choose $p_i\in [0,1]$ such that $p_{d+1} \ne 0$, $\sum_{i=1}^{d+1} p_i = 1$, and $p_i = 0$ iff $w_{ik}=0$ for all $k$.   We then put
\begin{eqnarray}
\alpha_i = {1\over{p_{d+1}}} \left(w_{ii} - p_i\right),  \qquad i=2,\ldots, d,
\end{eqnarray}
and thereby set
\begin{eqnarray}
\beta_{ij} &=& {1\over{p_{i}}} (w_{ij} - p_{d+1}\alpha_i),  \qquad i\ne j,\nonumber\\
\beta_{i1} &=& {1\over{p_{i}}} \biggl(w_{i1} - p_{d+1}\alpha_i\Bigl(1-\sum_{j=2}^{d} \alpha_i\Bigr)\biggr),\nonumber\\
\beta_{1j} &=& {1\over{p_{1}}} (w_{1j} - p_{d+1}),
\end{eqnarray}
for $i,j=2,\ldots, d$.
These coefficients allow us to obtain $W_C$ in (\ref{wexpansion}) from ({\ref{moritaexp}}) as required.

This shows that $W_C$ can generically be decomposed into $\mathrm{rank}(W_C)+1$ processes, with properly chosen  $\{ p_i \}$, and as such the processes are highly non-unique, as are the density operators which generically admit non-unique decompositions.   
Incidentally, we remark that, 
if we employ the singular value decomposition, we have the unique expansion
$W_C=\sum_i^d r_i \ket{u_i}\bra{v_i}$ with two bases $\mathcal{B}_u = \{\ket{u_i}\}$ and $\mathcal{B}_v = \{\ket{v_i}\}$ and $r_i \ge 0$ 
in terms of $d$ pairs of states at most.
However, the two bases $\mathcal{B}_u$ and $\mathcal{B}_v$ may have $\braket{u_i |v_i}=0$ for some $i$, in which case the two states
cannot appear as the initial and final states of a process ({\it i.e.}, the latter cannot be observed by postselection when the former is preselected), and accordingly they cannot specify a physical process.  Such unphysical cases are excluded in our expansion ({\ref{moritaexp}}).

\end{document}